\documentclass[5p]{elsarticle}
\usepackage{amsmath, amssymb, amsthm}
\usepackage{accents}
\usepackage[mathscr]{euscript}
\usepackage{mathrsfs}
\usepackage{mathtools}
\usepackage{textcomp, gensymb}
\usepackage{booktabs}
\usepackage{multirow}
\usepackage{array}
\usepackage{xcolor}
\usepackage[pdfauthor=author]{hyperref}
\usepackage[capitalize]{cleveref}
\usepackage[caption=false,font=footnotesize]{subfig}
\usepackage{comment}

\journal{Energy Conversion and Management}

\begin{document}

\begin{frontmatter}
\title{A Dynamically Similar Lab-Scale District Heating Network via Dimensional Analysis}

\cortext[cor1]{Corresponding author}
\address[a1]{Department of Mechanical and Aerospace Engineering, The Ohio State University, Columbus, 43210, 
OH,USA}

\author[a1]{Audrey Blizard\corref{cor1}}
\ead{blizard.1@osu.edu}

\author[a1]{Stephanie Stockar\corref{}}
\begin{abstract}
    Strict user demands and large variability in external disturbances, along with limited richness in the data collected on the daily operating conditions of district heating networks makes the design and testing of novel energy-reducing control algorithms for district heating networks challenging. This paper presents the development of a dynamically similar lab-scale district heating network that can be used as a test bench for such control algorithms. This test bench is developed using the Buckingham $\pi$ theorem to the match the lab-scale components to  the full-scale. By retaining the relative thermodynamics and fluid dynamics of a full-scale network in the lab-scale system, the experimental setup allows for repeatability of the experiments being performed and flexibility in the testing conditions. Moreover, the down-scaling of the experiment is leveraged to accelerate testing, allowing for the recreation of operating periods of weeks and months in hours and days. A PID controller is implemented on the lab-scale test bench to validate its response against literature data. Results show 63\% efficiency during heating operations compared to 70\% efficiency for a similar full-scale system, with comparable pressure losses across the system. 
\end{abstract}
\begin{keyword}
district heating network\sep thermal fluid systems\sep dynamic similitude\sep optimization\sep control verification
\end{keyword}

\end{frontmatter}
\section{Introduction}
Reducing energy consumption is one of the main tools available to combat climate change and lessen humans' impact on the environment. A large source of energy demand is the heating and cooling of buildings. For instance, in 2015 these processes account for 35\% of the energy consumed by buildings in the United States \cite{doeChapterIncreasingEfficiency2015}. Increasing the process efficiency will decrease the carbon footprint of commercial and residential buildings and reduce their operating costs.\par
In this context, district heating networks (DHNs) are a promising method to efficiently deliver heat to buildings in an urban environment. Rather than implementing individual heating systems in each building, DHNs centralize the process, relying on economies of scale to increase efficiency and reduce carbon emissions. Additionally, because the heat is generated at a centralized plant, DHNs also simplify the process of integrating of a variety of advanced technologies, including combined heating and power and bio-fuels, along with the use and storage of energy generated by intermittently-available renewable energy sources, such as solar \cite{liDistrictHeatingCooling2017}. While DHNs offer inherent performance benefit over traditional heating methods, the system performance can be greatly improved through advanced control strategies. Existing controllers rely on limited information, mainly the supply and return temperatures and the network pressure differential to set the supply temperature and initial flow rate \cite{vandermeulenControllingDistrictHeating2018}. While simple to implement, these controllers are ineffective in meeting the heat demands of users especially in large and highly interconnected DHNs due to the lack of granular data on the network's current operating conditions and limited knowledge of the network's future demands \cite{vandermeulenControllingDistrictHeating2018}. Conventional control strategies also lack the flexibility needed to take full advantage of inconsistent renewable energy sources \cite{xuQuantificationFlexibilityDistrict2020}. Optimized control algorithms enable existing DHNs to be operated more efficiently, reducing energy consumption and providing a more comfortable user experience at a lower operational cost. Preliminary studies of model-based control algorithms for DHNs, which are able to account for predicted demands and disturbances, have been shown to reduce the energy consumption by up 34\% \cite{delorenziSetupTestingSmart2020}.\par
However, in most cases, the validation of novel control algorithms for DHNs has been focused on tests performed in simulation environments \cite{giraudOptimalControlDistrict2017, vanhoudtActiveControlStrategy2018, delorenziPredictiveControlCombined2022}. Without validation of the control strategies on the physical system, it is difficult to guarantee that similar performance improvements observed in simulations will be obtained in real-world DHNs. There are some examples of physical tests benches for DHNs \cite{zhuSteadyStateBehavior2021,vanderheijdeDynamicEquationbasedThermohydraulic2017}, but none are sufficient for the testing of novel control algorithms due to their narrow focus on individual components of the DHN network. Additionally, there are examples of control algorithms being tested on real world DHNs \cite{daineseDevelopmentApplicationPredictive2019,hermansenModelPredictiveControl2022}. However, these tests required significant modifications to the communication infrastructure to enable the implementation this controller on an operational DHN. Furthermore, when performing tests on full-scale DHNs, there is unpredictability in the external disturbances, such as ambient temperature, solar irradiation, and occupancy times. The performance characteristics of DHNs are highly seasonal, and a validation test of new controllers across the variety of operating conditions experienced by DHNs can take weeks or even months. For example, one of the preliminary tests discussed above were conducted over an entire winter season \cite{daineseDevelopmentApplicationPredictive2019}.\par
A dynamically similar lab-scale test bench will eliminate challenges associated with the validation of novel control strategies, allowing for the repeatability in the experiments being performed, flexibility in the conditions being explored, and reducing the time needed to perform tests, while ensuring the applicability of the control algorithm to real-world DHNs. Existing literature supports the idea of using dynamically scaled experimental setups in the design of novel controllers, with the Buckingham $\pi$ Theorem being the most common method of obtaining dynamically similar setups. For example, this technique has been show effective in matching airborne wind energy systems dynamics \cite{cobbLabScaleExperimentalCharacterization2018} and vehicle dynamics during high speed maneuvers \cite{makarunTestingPredictiveVehicle2021}. The Buckingham $\pi$ Theorem is a formalization of the procedure of dimensional analysis which works by identifying the relevant relationships within a system and systematically scaling them while retaining their relative proportions \cite{buckinghamPhysicallySimilarSystems1914}. This technique is commonly used to create smaller and economical experimental test benches in a variety of fields, and its use in the dynamic scaling of thermo-fluid systems is well known. For example, the Buckingham $\pi$ Theorem has been applied in the analysis of the heat equation \cite{evansDimensionalAnalysisBuckingham1972} and more specifically, has been used in the design of the thermal characteristics of spacecrafts \cite{jonesThermalSimilitudeStudies1968}. The Buckingham $\pi$ theorem has been used in DHNs to analyze individual components of the network, but has never been used to simultaneously scale all components of the network. For example, a reduce parameter model of a building's temperature dynamics was obtained through the correlation of different heating and cooling modes into related nondimensional groups \cite{ciullaEvaluationBuildingHeating2017,chaichanaHeatLoadModelling2020}. Similarly, applications to the characterization of the heat transfer rate in heat exchangers was developed based on their geometry \cite{sukarnoNondimensionalAnalysisHeat2021, tarradCorrelationAirSideHeat2015}. A model of a DHN pipe network was created using the Buckingham $\pi$ Theorem with the goal of reducing the number of network characteristics needed to accurately predicting the heat losses \cite{carnogurskaMeasurementMathematicalModelling2016}. Finally, this approach has been used to perform a detailed analysis of the pressure losses of a liquid flow in a single length of pipe \cite{dasariCorrelationsPredictionPressure2014}. While the Buckingham $\pi$ Theorem is an established method for scaling the static responses of systems and analyzing individual components of DHNs, this technique has never been used in the scaling the dynamic response of an entire district heating network with the intended application of model validation and control testing. Using the Buckingham $\pi$ Theorem, this paper will present the design of a dynamically similar lab-scale test bench that will allow for rapid testing of new control algorithms on physical hardware, and the nondimensional framework developed in this paper provides a method to compare the results obtained on the scaled network to that of a full-sized DHN. As DHNs operate in both the thermodynamic and fluid domains, the characterization of both has to be considered in the scaling procedure.\par
The remainder of this paper is organized as follows. First, \cref{sec:network} provides a description of the components of the lab-scale DHN, along with the equations used to model the network's dynamics. \Cref{sec:nondim} describes the application of the Buckingham $\pi$ theorem to establish the nondimensionalization parameters used to characterize the system. Then, \cref{sec:values} presents ranges for the size of components in a full-scale DHN found in literature and describes the desired parameters for the equivalent lab-scale components using the established $\pi$ groups. In \cref{sec:validation}, the lab-scale system is validated against simulation results from an equivalent full-scale system. Finally, conclusions are presented in \cref{sec:conclusion}.

\section{District Heating Networks}
\label{sec:network}
\subsection{Experimental Setup}
The lab-scale setup considered in this paper is representative of a two-user DHN, however the sizing methodology presented is applicable to any network size. The two buildings in this network are represented by fluid-filled acrylic boxes. These thermal masses contain a submerged winding of copper pipe acting as the heat exchangers in a full-scale network to transfer heat from the distribution network into the buildings. Submerged impellers are used to distribute the heat throughout the thermal masses and ensure the fluid is well-mixed. \par
\begin{figure}
    \centering
    \includegraphics[width=3in]{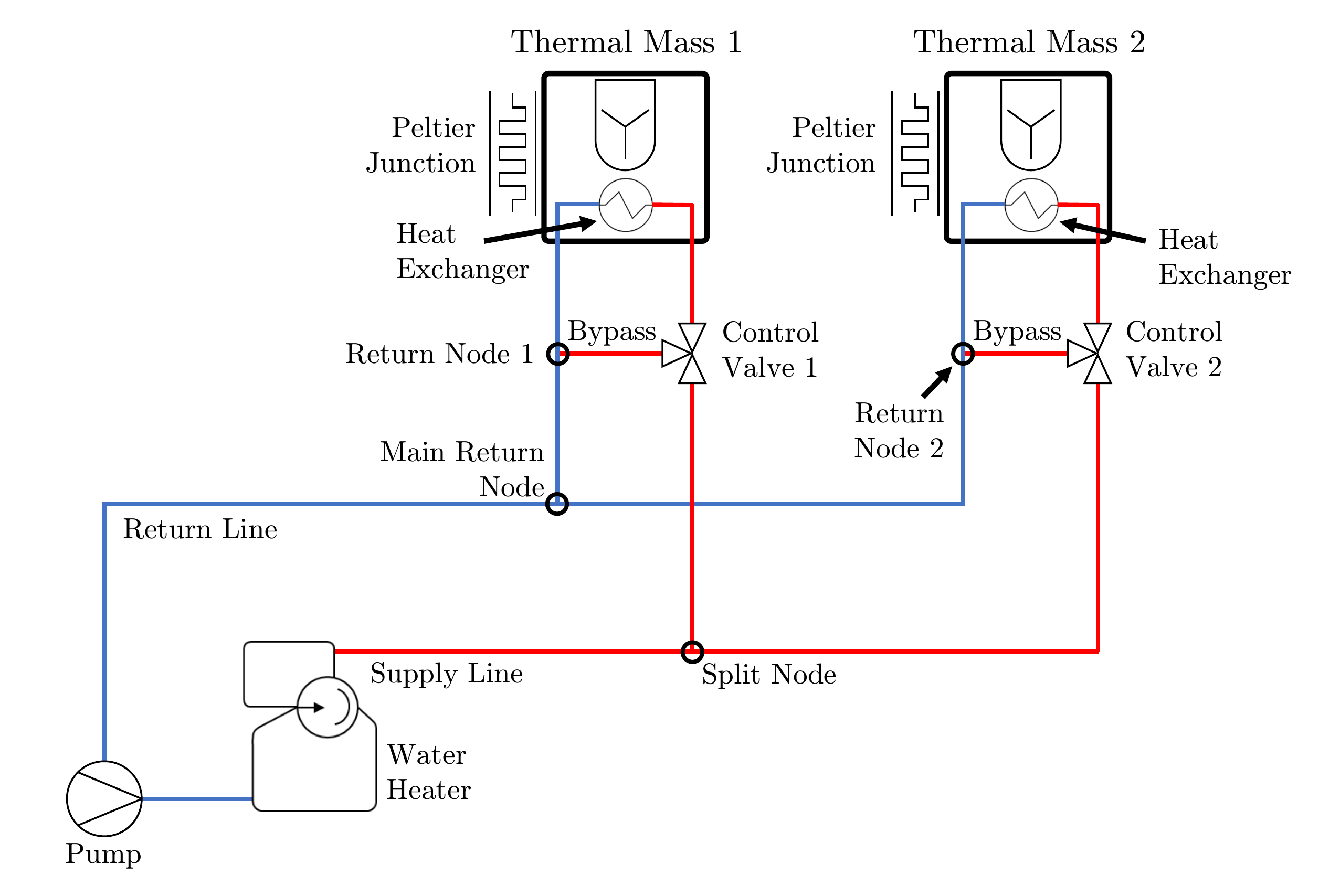}
    \caption{Layout of the two-user lab-scale DHN with components labeled.}
    \label{fig:layout}
\end{figure}
\begin{figure}
    \centering
    \includegraphics[width=3.25in]{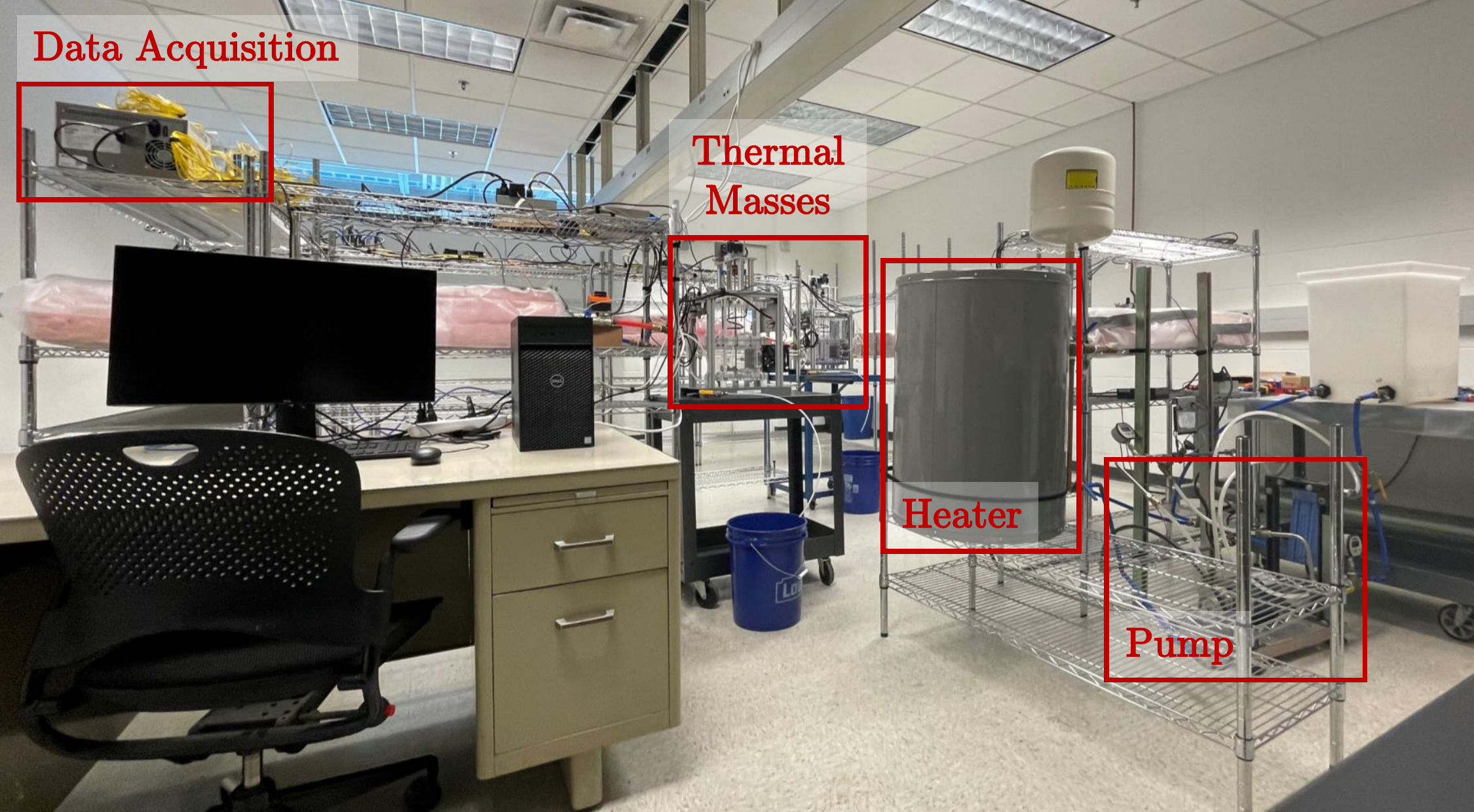}
    \caption{Photograph of experimental setup with relevant components labeled.}
    \label{fig:setup}
\end{figure}

The distribution network, made from 1/2" PEX pipes, circulates water from the heating plant to the thermal masses. This network consists of a main supply line that splits into two loops, one for each user. Each loop has a characterized control valve that divides the water between the user and a bypass. The user branch sends water to the heat exchanger, while the bypass branch diverts water from the heat exchanger. This division allows the operator to control the mass flow rate in the heat exchanger, adjusting the heat supplied to the thermal mass to control its temperature to a desired value, $T_{set}$. The flows from each heat exchanger and bypass are mixed at return nodes 1  and 2 and the user loops rejoin at the main return node. From here, the water is sent back to the centralized plant through the main return line to be reheated. The network is connected to a residential 30-gal water heater and a fixed-speed 3/4 HP pump, which reproduce the connection with the heating plant in a full-scale network. The detailed discussion of the experimental setup and its construction is presented in Krieger et al. \cite{kriegerDesignVerificationSmallScale2019}. A diagram of the described system is shown in \cref{fig:layout}, while a picture of the physical setup is shown in \cref{fig:setup}.\par
The component sizing in the original design were performed based on a static similarity analysis. However, an exact and systematic method to ensure dynamic similarity is needed to guarantee the performance of the lab-scale system is representative of a full-scale DHN.\par
\subsection{Dynamic System Model of a DHN}
The derivation of a physics-based model of a DHN, applicable to both the full-scale and lab-scale network is presented in this section. The derivation of these fundamental equations allows for the identification of the variables relevant to the system's behavior. Additionally, these equations will be used to simplify the dimensional analysis performed in \cref{sec:nondim} as these equations describe the relationship between the variables and parameters in the system.\par
The system is divided into three major components: the water supply, the distribution pipes, and the buildings, all of which exhibit both temperature and fluid dynamics. 
The lab-scale system is supplied by a water heater and pump that deliver water with a set supply temperature $T_s$ and initial mass flow rate $\dot{m}_I$ to the network. These values serve as the inlet values for the first pipe in the distribution network. The distribution network is then used to circulate this heated water to the buildings. Each pipe in the distribution network must be modeled in both the thermal and fluid domains. The bulk temperature $\left(T_p\right)$, which is the relevant temperature in each pipe segment, is modeled using the conservation of energy equation
\begin{equation}
\label{eq:Tpipe}
    \frac{dT_p}{dt} = \frac{\dot{m}}{\rho V}(T_{p_{in}}-T_p)-\frac{hA_s}{\rho c_p V}(T_p-T_{a})
\end{equation}
where ${\dot{m}}$ is the mass flow rate of water through the segment, $T_{a}$ is the ambient temperature, $\rho$ is the fluid density, $hA_s$ is the conductive heat transfer coefficient of the pipe segment, $c_p$ is the specific heat of water at constant pressure, assumed constant, and $V$ is the volume of each pipe segment. Finally,  $T_{p_{in}}$ is the inlet temperature obtained from the conservation of energy equation of the previous pipe. Note that for the supply line, $T_{p_{in}}=T_s$. The pipe's volume is given by
\begin{equation}
\label{eq:vol}
    V=\frac{\pi}{4}D^2l
\end{equation}
where $D$ is the pipes internal diameter and $l$ is the length of the pipe segment.\par
In the fluid domain, the pressure drop $\left(\Delta P\right)$ across each pipe segment is calculated using
\begin{equation}
    \label{eq:P}
    \Delta P = k_{tot}\left(\frac{\dot{m}}{A_c} \right)^2
\end{equation}
where $A_c$ is the pipe's cross sectional area and $k_{tot}$ is a pressure loss coefficient representative of both the distributed and concentrated pressure losses \cite{cimbala2006fluid}. 
In practice, DHNs are self balancing, meaning the mass flow $\left(\dot{m}\right)$ is divided between the branches to equalize the pressure losses throughout the network, resulting in a relative mass flow rate that is inversely proportional to $k_{tot}$. Hence the mass flow rate in each segment can be determined offline by solving the set of algebraic equations characterizing the pressure losses in each segment. The total pressure loss for each user loop $\Delta P_{Li},\ i = 1,2$ can be calculated as a function of mass flow rate by summing the $\Delta P$ values for each pipe segment. The pressure balance in the network can be enforced using
\begin{equation}
\label{eq:Pbal}
\Delta P_{L1}\left(\dot{m}_{1}\right) = \Delta P_{L2}\left(\dot{m}_{2}\right)
\end{equation}
 where $\dot{m}_i$ is the mass flow in each user loop. Conservation of mass ensures that
\begin{equation}
\label{eq:mdot}
\dot{m}_{1}+\dot{m}_{2} = \dot{m}_{I}
\end{equation}
Combining \cref{eq:Pbal} and \cref{eq:mdot}, the flow rate split between the branches can be calculated.\par
The temperature dynamics of the thermal mass is modeled using the conservation of energy equation, following the same principle as the pipe segments in \cref{eq:Tpipe}:
\begin{equation}
\label{eq:TThM}
    \left(\rho c_pV\right)_{ThM} \frac{dT_{ThM}}{dt}= \dot{Q}_{in} -\dot{Q}_{out}
\end{equation}
where $\left(\rho c_pV\right)_{ThM}$ is the heat capacity of the thermal mass, $\dot{Q}_{out}$ is the heat lost by the thermal mass to the environment, and $\dot{Q}_{in}$ is the heat provided to the thermal mass by the network through the heat exchanger.
The rate of heat transferred into the thermal mass is given by
\begin{equation}
\label{eq:QinThM}
    \dot{Q}_{in} = \left( hA_s\right)_{HX}\left(T_{HX}-T_{ThM}\right)
\end{equation}
where $T_{HX}$ is the bulk temperature of the water in the heat exchanger, and $(hA_s)_{HX}$ is the convective heat transfer coefficient of the heat exchanger. Similarly, the heat transferred from the thermal mass to the environment is
\begin{equation}
\label{eq:QoutThM}
    \dot{Q}_{out} = \left( hA_s\right)_{ThM}\left(T_{ThM}-T_{a}\right)
\end{equation}
where $(hA_s)_{HX}$ is the convective heat transfer coefficient of the thermal mass.\par
\Cref{eq:Tpipe,eq:vol,eq:P,eq:Pbal,eq:mdot,eq:TThM,eq:QinThM,eq:QoutThM} provide a model of the network that will serve as the foundation to perform the scaling of the system. Having this model simplifies the process of creating a similar lab-scale system by proving an understanding of the underlying relationship between the relevant variables.
\section{Nondimensionalization of System Equations}
\label{sec:nondim}
To ensure applicability of tests performed on the lab-scale system, the transient and steady state responses of the full-scale system must be scaled to the lab-scale. The multi-domain nature of DHNs requires the lab-scale model to match both the heat transfer characteristics and the fluid dynamics of the original system. This section presents the extension of the  Buckingham $\pi$ Theorem for the uniformly scaling of all components of a full-scale DHN. The procedure established by the Buckingham $\pi$ Theorem is followed to identify the relevant nondimensionalization parameters for the system and use them to create dimensionless $\pi$ groups that describe the behavior of the system. Finally, from these $\pi$ groups, the nondimensional form of the system equations is found.\par
The first step of the Buckingham $\pi$ Theorem is to identify the variables that are relevant to the system. From the equations presented in \cref{sec:network}, the 20 relevant variables associated with the system are
\begin{equation}
f\left(
\begin{gathered}
\label{eq:vars}
t,\ T_s,\ T_a,\ T_p,\ T_{HX},\ T_{ThM},\ \dot{m}_I,\ \dot{m},\ \rho,\\
c_p,\ D,\ L,\ hAs,\ k_{tot},\ A_c,\ \rho_{ThM},\\
c_{p_{ThM}},\ V_{ThM},\ hAs_{ThM},\ hAs_{HX} 
\end{gathered}
\right) = 0
\end{equation}\par
However, due to design constraints, the available sizes of some components, and lab space restrictions, there are a limited number of modifiable design parameters. Due to the aforementioned constraints, matching all 20 variables for every component of the system is impossible. Instead, using the system model, it is possible to combine some of the variables into groups. Because the equations governing the system's dynamic responses and losses have been established, the relationship between the some of the variables and these variables can be grouped together during the nondimensionalization process. This will reduce the number of dimensionless quantities to be matched, while still providing a representative description of the system. Leveraging this information, the variables listed in \cref{eq:vars} are reduced to the following
\begin{equation}
f\left(
\begin{gathered}
\label{eq:varsgroup}
t,\ T_s,\ T_p,\ T_{HX},\ T_{ThM},\ \dot{m}_I,\\
D,\ \rho,\ \frac{\dot{m}}{\rho V},\ \frac{hA_s}{\rho c_pV}\left(T_p-T_{a}\right),\\
\Delta P,\ \left(\rho c_pV\right)_{ThM},\ \ \dot{Q}_{in},\ \dot{Q}_{out}
\end{gathered}
\right)=0
\end{equation}\par
The next step in the Buckingham $\pi$ Theorem is to identify the fundamental units of the variables in the system. Dimensional formulas can be used to show the fundamental units associated with each physical quantity and to determine what terms are needed to generate each nondimensional group. 
The dimensional formulas for each variable group in the reduced list are
\begin{subequations}
\label{eq:dims}
\begin{equation}
    t = [t]
\end{equation}
\begin{equation}
    T_s,T_p,T_{HX},T_{ThM} = [T]
\end{equation}
\begin{equation}
    \dot{m}_I = \left[Mt^{-1}\right]
\end{equation}
\begin{equation}
    D = [L]
\end{equation}
\begin{equation}
    \rho = \left[ML^{-3}\right]
\end{equation}
\begin{equation}
    \frac{\dot{m}}{\rho V} = \left[t^{-1}\right]
\end{equation}
\begin{equation}
    \frac{hA_s}{\rho c_pV}(T_p-T_{a}) = \left[Tt^{-1}\right]
\end{equation}
\begin{equation}
    \Delta P=\left[ML^{-1}t^{-2}\right]
\end{equation}
\begin{equation}
    \left(\rho c_pV\right)_{ThM}=\left[ML^2t^{-2}T^{-1}\right]
\end{equation}
\begin{equation}
    \dot{Q}_{in},\dot{Q}_{out}=\left[ML^2t^{-3}\right]
\end{equation}
\end{subequations}\par
From these dimensional formulas, it can be seen that four fundamental units appear in the variables used to describe the relevant system dynamics. These units are length $\left(L\right)$, time $\left(t\right)$, mass $\left(M\right)$, and temperature $\left(T\right)$. Therefore, four fundamental quantities are needed to create the independent, nondimensional variable groups.
The quantities selected are fluid density, initial mass flow rate, supply temperature, and internal pipe diameter and are summarized in \cref{table:NondimParams} . These values are selected because first, they best characterize the scale of operation for each component in the system, and second, are the ones subject to the most challenging constraints (dimensions, specifications) in the lab-scale design.
\begin{table}
\caption{Nondimensionalization parameters selected to create the $\pi$ groups}
\centering
\begin{tabular}{lcc}
\toprule
 \bf{Parameter} & \bf{Symbol} & \bf{ Unit}\\
\midrule
 Density of operating fluid & $\rho$ & $ML^{-3}$ \\
 Initial mass flow rate & $\dot{m}_{I}$ & $Mt^{-1}$\\  
 Supply temperature & $T_{s}$ & $T$\\
 Pipe internal diameter & $D$ & $L$ \\
\bottomrule
\end{tabular}
\label{table:NondimParams}
\end{table}
From the above steps, it can be seen there are fourteen relevant variable groups and four fundamental units. Therefore, according to the Buckingham $\pi$ Theorem, a total of ten independent $\pi$ groups can be found. The process of finding the $\pi$ group for a single variable is
\begin{enumerate}
    \item Identify the fundamental units of the variable using its dimensional formula, given in \cref{eq:dims}.
    \item Note the exponent for each of the fundamental units.
    \item Eliminate the units of the variable using the appropriate power of the  nondimensionalization parameters (\cref{table:NondimParams}) in order of $T_s$, $\dot{m}_I$, $\rho$, and $D$.
    \item Write the final nondimensional group from the original variable and nondimensionalization parameters raised to the appropriate powers.
\end{enumerate}
This procedure is followed to nondimensionalize the variables listed in \cref{eq:varsgroup}. The resulting $\pi$ groups are listed below.\par
Nondimensional time is found using 
\begin{equation}
\label{eq:tstar}
    t^* = t\cdot\frac{\dot{m}_I}{\rho D^3}
\end{equation}\par
The nondimensional temperatures of each pipe segment in the network can be found by dividing the current temperature by the supply temperature.
\begin{equation}
\label{eq:Tstar}
    T_p^* = \frac{T_p}{T_s},\quad T_{HX}^* = \frac{T_{HX}}{T_s}
 \end{equation}\par
Because the temperature of the fluid in the thermal mass can be modulated by the bypass valve, it is independent from the network supply temperature $T_s$, and the desired setpoint, $T_{set}$, can be decided arbitrarily, provided that $T_{set}<T_{s}$. Therefore, the current thermal mass temperature must be normalized by subtracting the setpoint before dividing by the supply temperature
 \begin{equation}
\label{eq:Tthmstar}
    T_{ThM}^*=\frac{T_{ThM}-T_{set}}{T_s}
 \end{equation}\par
There are three $\pi$ groups needed to characterized the pipes, denoted as $\pi_1-\pi_3$. The first two groups, $\pi_1$ and $\pi_2$ are from conservation of energy equation for the pipe \cref{eq:Tpipe}, where $\pi_1$ is the nondimensional coefficient for the heat supplied into the pipe
\begin{equation}
\label{eq:pi1}
\pi_{1} = \frac{\dot{m}}{\rho V}\cdot\frac{D^3\rho}{\dot{m}_I}
\end{equation}
and $\pi_2$ comes from the term used to describe the heat lost to the environment by the pipe.
\begin{equation}
\label{eq:pi2}
\pi_{2}=\frac{hA_s}{\rho c_pV}(T_p-T_{a})\cdot\frac{D^3\rho}{\dot{m}_IT_s}
\end{equation}
The final group for the pipes, $\pi_3$, comes from the fluid dynamics model of the pipe \cref{eq:P} and is used to ensure the pressure losses are consistent between the network scales.
\begin{equation}
\label{eq:pi3}
\pi_3 =\Delta P\cdot\frac{\rho D^4}{\dot{m}^2}
\end{equation}\par
The dynamics of the building components, given by \cref{eq:TThM} can be described using $\pi_4-\pi_6$, where $\pi_4$ is the nondimensional heat capacity of the building
\begin{equation}
\label{eq:pi4}
\pi_{4} =(\rho c_pV)_{ThM}\cdot\frac{\rho T_sD}{\dot{m}^2}
\end{equation}
The second group for the building dynamics, $\pi_5$, describes the energy transferred into the thermal mass from the heat exchanger
\begin{equation}
\label{eq:pi5}
\pi_{5}=\dot{Q}_{in}\frac{\rho^2D^4}{\dot{m}^3}
\end{equation}
and the last group $\pi_6$ describes the heat lost by the thermal mass to the environment
\begin{equation}
\label{eq:pi6}
\pi_6 = \dot{Q}_{out}\frac{\rho^2D^4}{\dot{m}^3}
\end{equation}\par
Using the $\pi$ groups, the system equations are then written in their dimensionless forms. Having the nondimensional form of these equations allows for the modeling of the network behavior in the nondimensional space, which makes comparison between the full-scale and lab-scale systems possible.
The temperature dynamics equation in the pipes in \Cref{eq:Tpipe} is reformulated as
\begin{equation}
    \label{eq:Tpipend}
    \frac{dT_p^*}{dt^*} = \pi_1\left(T_{p_{in}}^*-T_{p}^*\right)-\pi_2
\end{equation}
while the temperature dynamics equation for the thermal masses, \cref{eq:TThM}, is rewritten as 
\begin{equation}
    \label{eq:TThMnd}
    \pi_4\frac{dT_{ThM}^*}{dt^*}=\pi_5-\pi_6
\end{equation}

\section{Selection of the Lab-Scale System Parameters}

\label{sec:values}
\begin{table*}
\caption{Average network component values for full-scale and lab-scale systems}
\label{table:val}
\centering
\begin{tabular}{lcccccc}
\toprule
\multicolumn{5}{l}{\bf{Nondimensionalization Parameters}}\\
\midrule
&&\multicolumn{2}{c}{Full-Scale}&\multicolumn{2}{c}{Lab-Scale}&\\
Parameter& Symbol & Value & Unit & Value & Unit & Source\\
\midrule
Density of network fluid & $\rho$ & 971&$\left[kg/m^3\right]$ &994& $\left[kg/m^3\right]$ &\cite{cimbala2006fluid}\\
Initial mass flow rate& $\dot{m}_I$ & 20& $\left[kg/s\right]$& 0.0862& $\left[kg/s\right]$ & \cite{anconaApplicationDifferentModeling2019, salettiDevelopmentAnalysisApplication2020}\\
Supply temperature&$T_s$ &80&$\left[C\right]$&36&$\left[C\right]$&\cite{anconaApplicationDifferentModeling2019, gabrielaitieneModellingTemperatureDynamics2007}\\
Pipe internal diameter&$\left[D\right]$&0.1&$\left[m\right]$&12&$\left[mm\right]$& \cite{gabrielaitieneModellingTemperatureDynamics2007, salettiDevelopmentAnalysisApplication2020}\\
\midrule[\heavyrulewidth]
\multicolumn{5}{l}{\bf{Pipe Characteristics}}\\
\midrule
&&\multicolumn{2}{c}{Full-Scale}&\multicolumn{2}{c}{Lab-Scale}&\\
Parameter& Symbol & Value & Unit & Value & Unit & Source\\
\midrule
Length of segment&$l$ & $20-100$ &$\left[m\right]$ & $2.5-11$&$\left[m\right]$& \cite{anconaApplicationDifferentModeling2019,piroutiEnergyConsumptionEconomic2013} \\
Pressure loss across segment&$\Delta P$ &0.01-0.2& $\left[MPa\right]$ & 1-10 & $\left[kPa\right]$&\cite{piroutiEnergyConsumptionEconomic2013,dallarosaLowenergyDistrictHeating2011}\\
 Conductive heat transfer coefficient&$hA_s$&5-90&$\left[W/K\right]$&0.23-1.0&$\left[W/K\right]$&\cite{masatinEvaluationFactorDistrict2016}\\
\midrule[\heavyrulewidth]
 \multicolumn{5}{l}{\bf{Thermal Mass Characteristics}}\\
 \midrule
 &&\multicolumn{2}{c}{Full-Scale}&\multicolumn{2}{c}{Lab-Scale}&\\
Parameter& Symbol & Value & Unit & Value & Unit & Source\\
\midrule
Heat capacity& $\left({\rho c_pV}\right)_{ThM}$ & $0.15-7$& $\left[GJ/K\right]$& $30-45$& $\left[kJ/K\right]$ & \cite{gambarottaDevelopmentModelbasedPredictive2019,ciullaEvaluationBuildingHeating2017, vanhoudtActiveControlStrategy2018} \\
 Pressure loss across heat exchanger&$\Delta P_{HX}$  & 0.03& $\left[MPa\right]$&2-3& $\left[kPa\right]$&\cite{piroutiEnergyConsumptionEconomic2013}\\
 Convective heat transfer coefficient&\multirow{2}{*}{$\left(hA_s\right)_{HX}$} &\multirow{2}{*}{5-12}&\multirow{2}{*}{$\left[kW/K\right]$}& \multirow{2}{*}{14.5-16}& \multirow{2}{*}{$\left[W/K\right]$}& \multirow{2}{*}{\cite{cadauModelintheLoopApplicationPredictive2018, bojicLinearProgrammingOptimization2000}}\\
of heat exchanger&&&&\\
Rate of heat lost by building& $\dot{Q}_{out}$ &70-230&$\left[kW\right]$&0-79&$\left[W\right]$&\cite{gambarottaDevelopmentModelbasedPredictive2019, salettiDevelopmentAnalysisApplication2020}\\
Desired internal temperature of building &$T_{Set}$&20&$\left[C\right]$&26&$\left[C\right]$& \cite{cadauModelintheLoopApplicationPredictive2018}\\
\bottomrule
\end{tabular}
\end{table*}
The physical parameters of the lab-scale system are selected to match the nondimensional large-scale values for all the previously established $\pi$ groups. 
A literature review was conducted to find representative dimensional values for the sizes of each component for a variety of real-world DHN configurations. This information is then used to calculate the values of the full-scale $\pi$ groups. The results of this literature review is summarized in \cref{table:val}.\par
Then, the unmodifiable components of the lab-scale system were considered. Specifically, the lab-scale network is supplied by a a fixed speed rotary vane pump and 30 gallon residential water heater. These two components constrained two of the
nondimensionalization parameters, namely the size of the pump determines $\dot{m}_I$, and the recovery rate of the water heater (the amount of water that can be heated during a one hour period) dictates $T_s$. Additionally, the network uses 1/2" PEX pipes for the supply network, fixing the value for $D$. By modifying the original experimental setup, the rest of the components in the lab-scale DHN are resized to ensure agreement between the full-scale and lab-scale $\pi$ groups. The ranges of the lab-scale values that result from matching the $\pi$ groups are presented in \cref{table:val}.\par
The lengths of the pipes are chosen to ensure agreement between the $\pi_1$ values. The pipes have been encased in two layers of R13 fiberglass insulation to reduce the heat lost by the pipes to the environment, matching the $\pi_2$ values. Additionally, the predicted pressure drops in the lab-scale network are calculated based on the friction, length, and geometry changes in the network and were found to be in the desired range to match the values of $\pi_3$ for the lengths selected.\par
\begin{figure}
    \centering
    \includegraphics[width=2.5in]{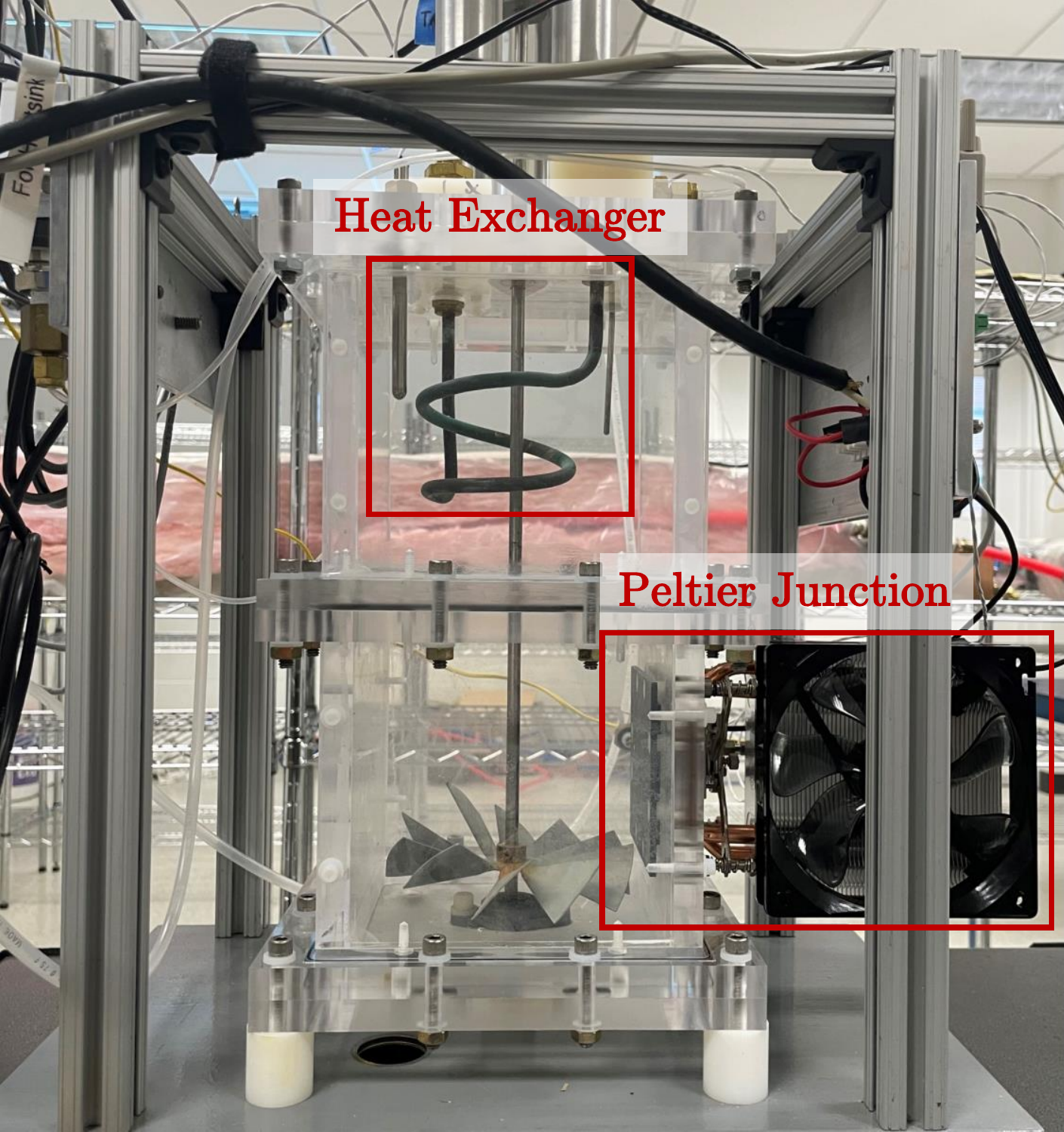}
    \caption{Photograph of thermal mass with embedded Peltier junction.}
    \label{fig:setupThM}
\end{figure}

The thermal mass characteristics are also matched between the full-scale and lab-scale. To decrease the volume required to match heat capacity of the thermal mass and increase the rate of heat transfer from the heat exchanger, the lab-scale thermal masses representing the buildings are filled with water, rather than the air as would be in full-scale system. The volume of the thermal masses is then chosen to match $\pi_{4}$. This results in the chosen volumes of $7000\ cm^3$ and $10400\ cm^3$ respectively. The desired steady-state temperature in the thermal mass is set to match the rate of heat extracted from the network $\left(\pi_5\right)$ during steady state operation. The desired building temperature is of particular relevance because this variable is used to modulate the bypass valves supplying the heat exchangers to meet the desired setpoint when each building is occupied. Finally, because the lab-scale DHN is indoors, the difference between the set point temperature $\left(T_{ThM}\right)$ and the ambient temperature $\left(T_{a}\right)$ is less than for an equivalent outdoor system. Hence, the required rate of heat transfer out of the buildings can not be met through natural convection alone. Additionally, the ambient temperature can vary widely throughout a day, and the ability to recreate these fluctuations is key to performing realistic tests. To address these issues, Peltier junctions are added to a wall of the thermal masses. An image of one of the thermal mass with the embedded Peltier junction can be seen in \cref{fig:setupThM}. These Peltier junctions remove heat from the thermal mass when a voltage is applied via the thermoelectric cooling effect. The addition of the Peltier junction to the thermal mass allows for the control of the rate of heat removed from each thermal mass, allowing for more flexibility in the tests that can be performed. The rate of heat lost by the thermal mass (\cref{eq:QoutThM}) must be modified to include the effects of the Peltier junction
\begin{equation}
\label{eq:Qoutp}
    \dot{Q}_{out} = \left(hA_s\right)_{ThM}\left(T_{ThM}-T_{a}\right)+\dot{Q}_{pelt}
\end{equation}
where $\dot{Q}_{pelt}$ is the rate of heat removed by the Peltier junction, which is set directly by modulating the power supplied to the Peltier junctions. For example, to replicate an ambient temperature of -5 C in the full-scale, the Peltier junctions must operate at approximately 40 W. The rate of heat loss caused by the Peltier junctions can be converted to a simulated ambient temperature $\left(T_{a_{sim}}\right)$ as
\begin{equation}
\label{eq:Tambsim}
T_{a_{sim}} = -\frac{hA_{s_{act}}}{hA_{s_{sim}}}\left(T_{ThM}-T_{a}\right)-\frac{\dot{Q}_{pelt}}{hA_{s_{sim}}}+T_{ThM}
\end{equation}
where $hA_{s_{act}}$ is the thermal mass's heat transfer coefficient and $hA_{s_{sim}}$ is the simulated heat transfer coefficient. The thermal masses' heat transfer coefficients $hA_{s_{sim}}$ and the ambient temperature, both components of $\pi_5$, are modulated by changing voltage supplied to the Peltier junction to match the full-scale values.
\section{Validation of the Resulting Lab-Scale DHN}
\label{sec:validation}
There is a limited amount of data available in literature on daily operating conditions of real-world DHNs at the individual user level. More often, models are developed for design and energy characterization purposes. For this reason, in the paper, simulation results are used to validate the similarity between the lab-scale and full-scale systems. 

\subsection{Data Acquisition System}
\begin{figure}
    \centering
    \includegraphics[width = 3in]{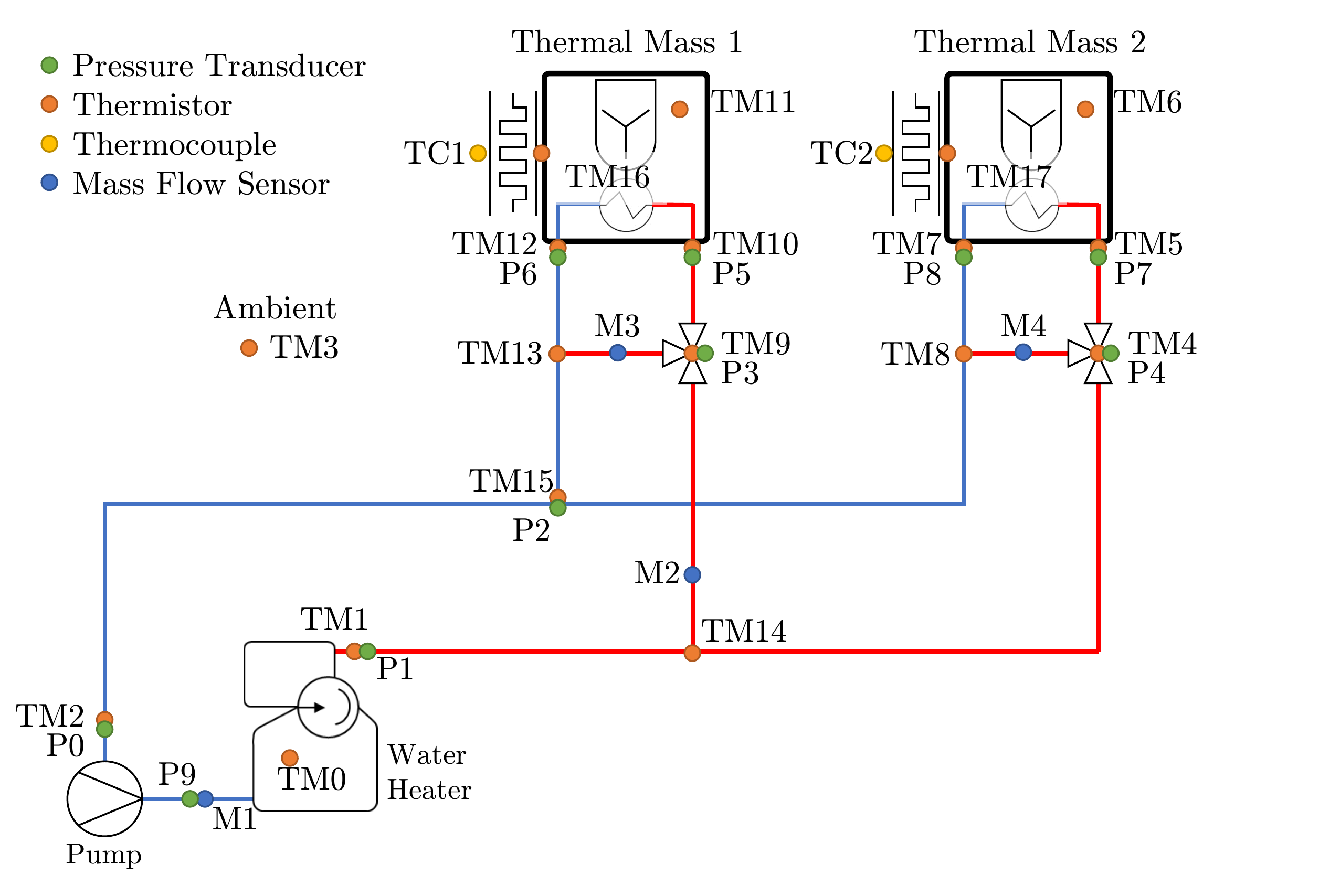}
    \caption{Diagram of the two-user lab-scale DHN with sensor locations labeled.}
    \label{fig:sensors}
\end{figure}

The experimental setup is outfitted with temperature, pressure and mass flow sensors to ensure all relevant dynamics are captured during operation. The system has two separate but synchronized data acquisition systems (DAQs). The first DAQ consists of three 8-channel USB data acquisition modules and is used for collecting temperature data from the 17 thermistors located throughout the network. The second system is a National Instruments PXIe1073 system, with a PXIe4353 and a PXIe6363 installed, which is used for temperature, pressure and flow rate data. The PXIe4353 provides thermocouple input channels, which are used to measure the cold-side temperature of the Peltier junctions. The PXIe6363 has 16 analog differential inputs, nine of which are used to collect data from the pressure transducers located throughout the network. Additionally, four analog input channels are used to collect data from the mass flow sensors. The PXIe6363 also has 4 analog outputs, which supply voltage to the two characterized control valves to modulate their positions during the experiment. The location of the sensors in the network are shown in \cref{fig:sensors}. NI LabVIEW is used to interface with both DAQs to synchronize and record the data, set the control valve positions, and display results in real time. 
\subsection{Description of Experiment}
Through the $\pi$ groups framework, any set of testing conditions can be recreated in the lab. In this paper, the response of the lab-scale system is evaluated against the simulation results presented in the paper by Saletti et al. \cite{salettiDevelopmentAnalysisApplication2020}. This test is representative of a typical day during winter operation and shows a diverse response range for the two thermal mases. In the full-scale experiment, a PID control algorithm was designed to act as a baseline for comparison to a novel control algorithm. This PID controller was implemented in a Model-in-the-Loop simulation to control two school buildings to a desired set temperature during their hours of operation. In the full-scale simulation, the maximum available supply mass flow rate for the two buildings was 20 kg/s, while the maximum supply temperature was 80 C, consistent with the ranges presented in \cref{sec:values}.\par 
A continuous 48 hour subset of the week-long simulation was selected to be emulated in the lab-scale experiment. In the full-scale experiment, thermal mass 1, associated with the smaller building (a school), started at its heated temperature and was immediately cut off from the heat supply and allowed to cool for 12 hours, before being heated again for an additional 12 hours. The occupancy times of thermal mass 2, representative of the larger building (a sports hall), were offset from that of thermal mass 1. The larger thermal mass remained at its heated temperature for two hours and 45 minutes before being set to cool for 11 hours. After its cooling period, thermal mass 2 was reheated for the remaining 10 hours and 15 minutes. These occupancy cycles were repeated twice during the 48 hour period. For the lab-scale experiment, the timings of cycles were scaled using \cref{eq:tstar}, resulting in the entire 48 hour period being recreated in just under 18 hours, a 63\% reduction in time needed to run the experiment.
To recreate the thermal masses' temperature profiles, two PID controllers are implemented in the lab-scale network, where the difference between the current thermal masses' temperatures and their desired setpoints are used to set the position of the bypass valves to drive the thermal masses to the desired temperature. During hours of operation, the setpoint was 28 C, while during cooling periods, the PID control setpoint was 0 C.\par
The ambient temperature profile used in the original simulation is show in \cref{fig:temp}. This ambient temperature was recreated in the lab-scale experiment using the Peltier junction. The power setpoint for each Peltier junction was controlled using
\begin{multline}
    \dot{Q}_{pelt} = \left(hA_s\right)_{ThM}\bigl(\left(T_{a}-T_{set}\right)_{Lab}\\ 
    - k_T\left(T_a-T_{set}\right)_{Full}\bigr)
\end{multline}
where $k_T$ is the ratio of nondimensionalization parameters for the full-scale and lab-scale temperatures:
\begin{equation}
    k_T = \frac{T_{s_{Lab}}}{T_{s_{Full}}}
\end{equation}
Each Peltier junction has a built-in PID controller that is used to track a desired power setpoint.
\subsection{Results}
\begin{figure}
    \centering
        \subfloat[Response of thermal mass 1 for full and lab-scale systems.]{\includegraphics[width=3in]{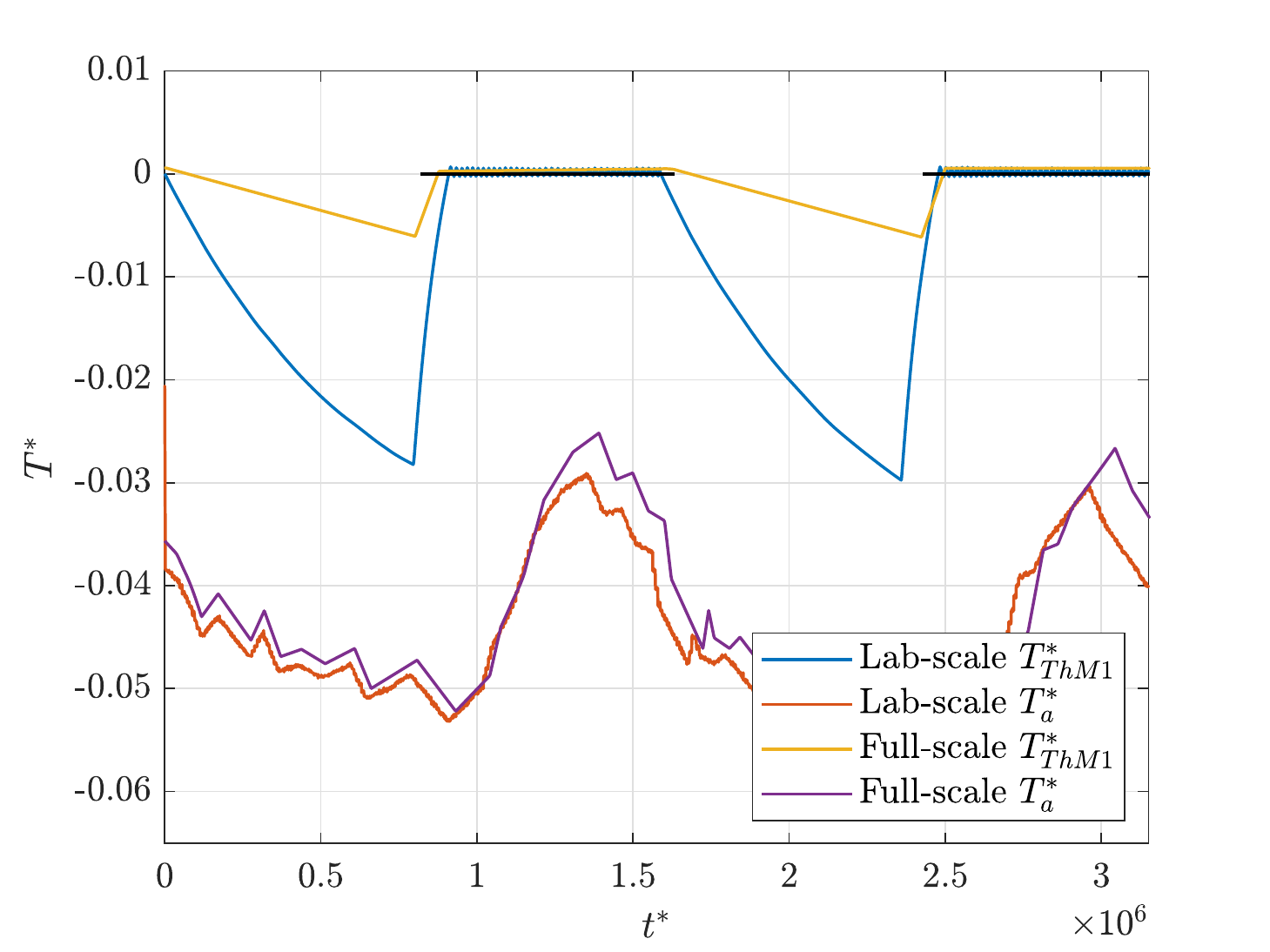}\label{fig:temp1}}
        \hfil
        \subfloat[Response of thermal mass 2 for full and lab-scale systems.]{%
        \includegraphics[width=3in]{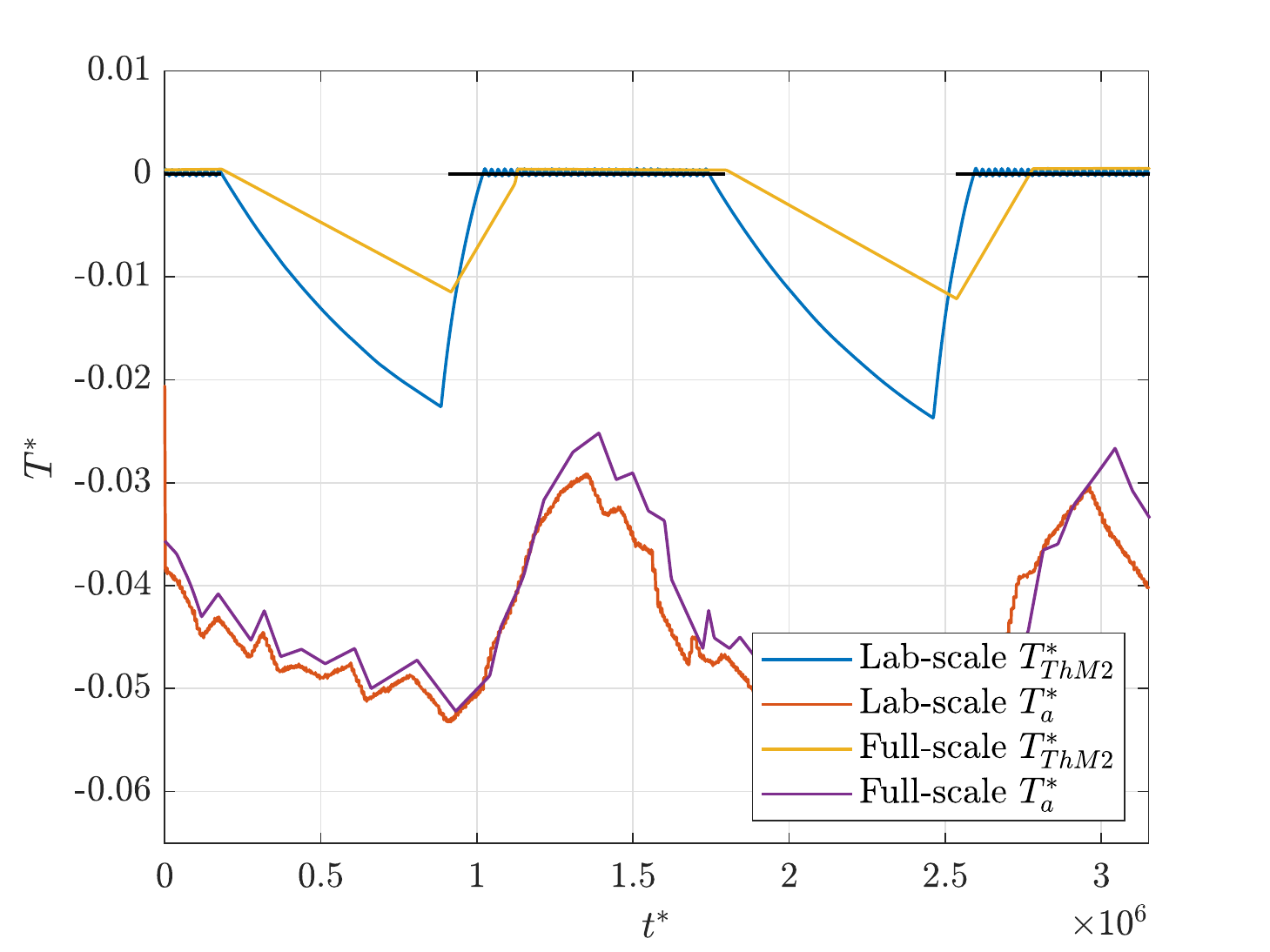}\label{fig:temp2}}
    \caption{Comparison between the lab-scale and full-scale systems.}
    \label{fig:temp}
\end{figure}

\begin{figure}
    \centering
    \includegraphics[width = 3in]{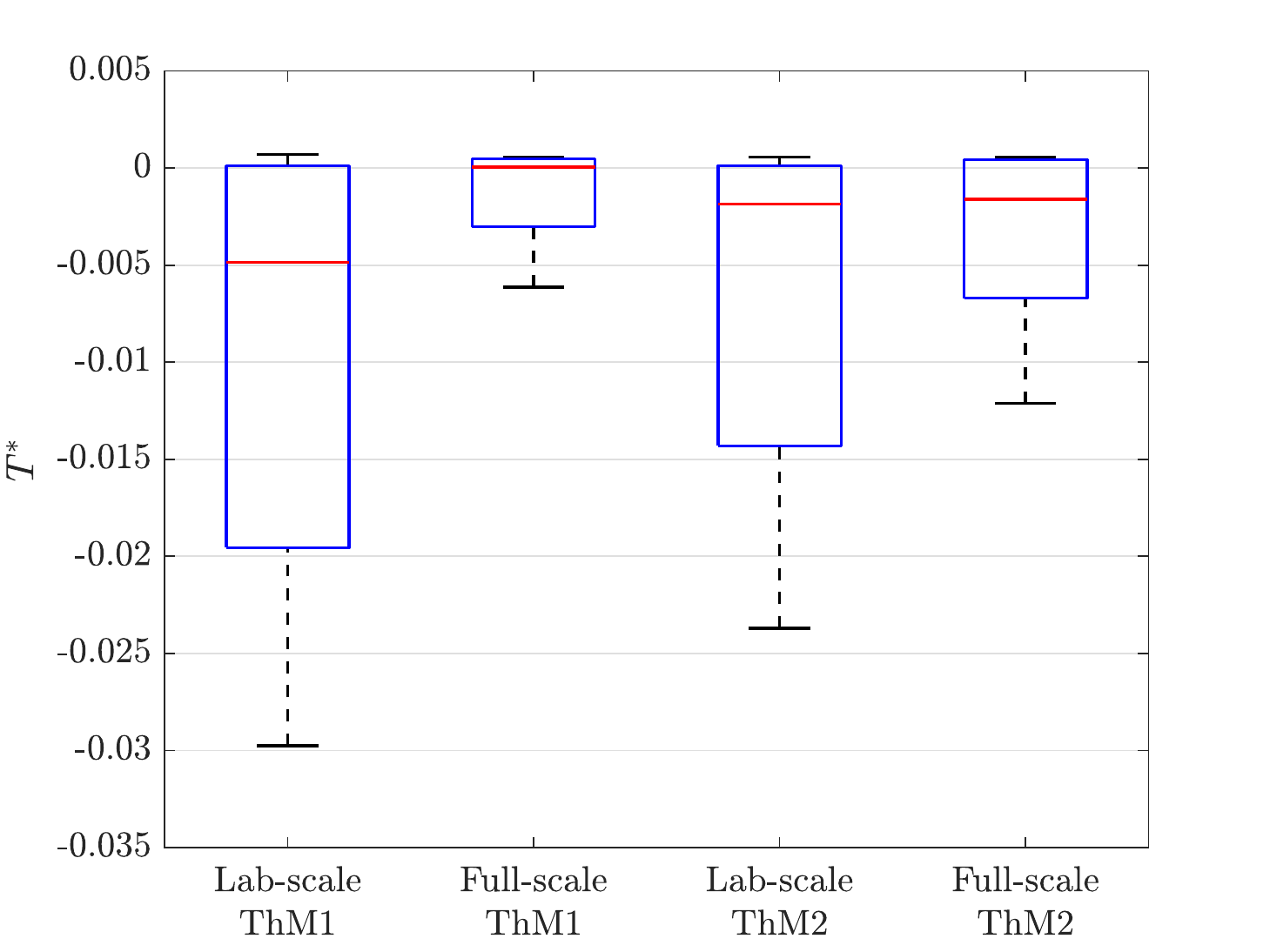}
    \caption{Median nondimensional thermal mass temperature with upper and lower quartiles for full-scale and lab-scale thermal masses.}
    \label{fig:error}
\end{figure}

The temperature response of the thermal masses are shown in \cref{fig:temp,fig:error}, specifically the nondimensional temperature of the thermal masses, along with the nondimensional simulated ambient temperatures for both the full-scale and lab-scale systems are shown in \cref{fig:temp1,fig:temp2} for thermal mass 1 and thermal mass 2, respectively. The result of the direct comparison shows a consistent dynamic response between the full-scale and the lab-scale system. Additionally, it can be seen that the simulated ambient temperature is able to effectively follow the desired full-scale temperature, with the RMS error being 0.22\% for both thermal masses.\par
\Cref{fig:error} shows the median nondimensional full-scale and lab-scale thermal mass temperature along with the limits of 25th and 75th percentiles for the temperature of both thermal masses. The mean and standard deviations of the temperatures are listed in \cref{table:error}, along with the ratio of the mean lab-scale temperatures to the mean full-scale temperatures. Consistently, the lab-scale results have a larger spread in the nondimensional thermal mass temperature, which is more evident in the smaller thermal mass ($ThM1$).\par
While the general trend between the full-scale and lab-scale thermal mass temperatures are similar, there is a large variation between the extreme values observed. This deviation can be attributed to the difference in values of $\pi_4$ between the particular buildings being represented. Specifically, $\pi_4$ describes the total heat capacity of the thermal mass, which is mainly effected by the building's volume. In the full-scale configuration, the nondimensional heat capacities were approximately 8.6 times bigger for both thermal mass 1 and thermal mass 2 in the full-scale simulation. These values are consistent with the ratio of the mean temperature values between the two scales. Moreover, when considering the differences in the heat transfer coefficients for different sized buildings, this provides an explanation of the variations observed. There was a large range of potential building volumes presented in literature; in the lab-scale setup, the thermal masses are sized more similarly to residential buildings, while the full-scale experiment used commercial/industrial sized buildings. The effects of various building sizes can be seen in Gambarotta et al. \cite{gambarottaLibrarySimulationSmart2017}, where the temperature dynamics of 12 different buildings are simulated in one network. The building temperatures in the lab-scale system will exhibit more rapid dynamic responses as compared to larger buildings, but will still be valid as a proving ground for new modeling and control techniques.\par
\begin{figure}
    \centering
    \includegraphics[width=3in]{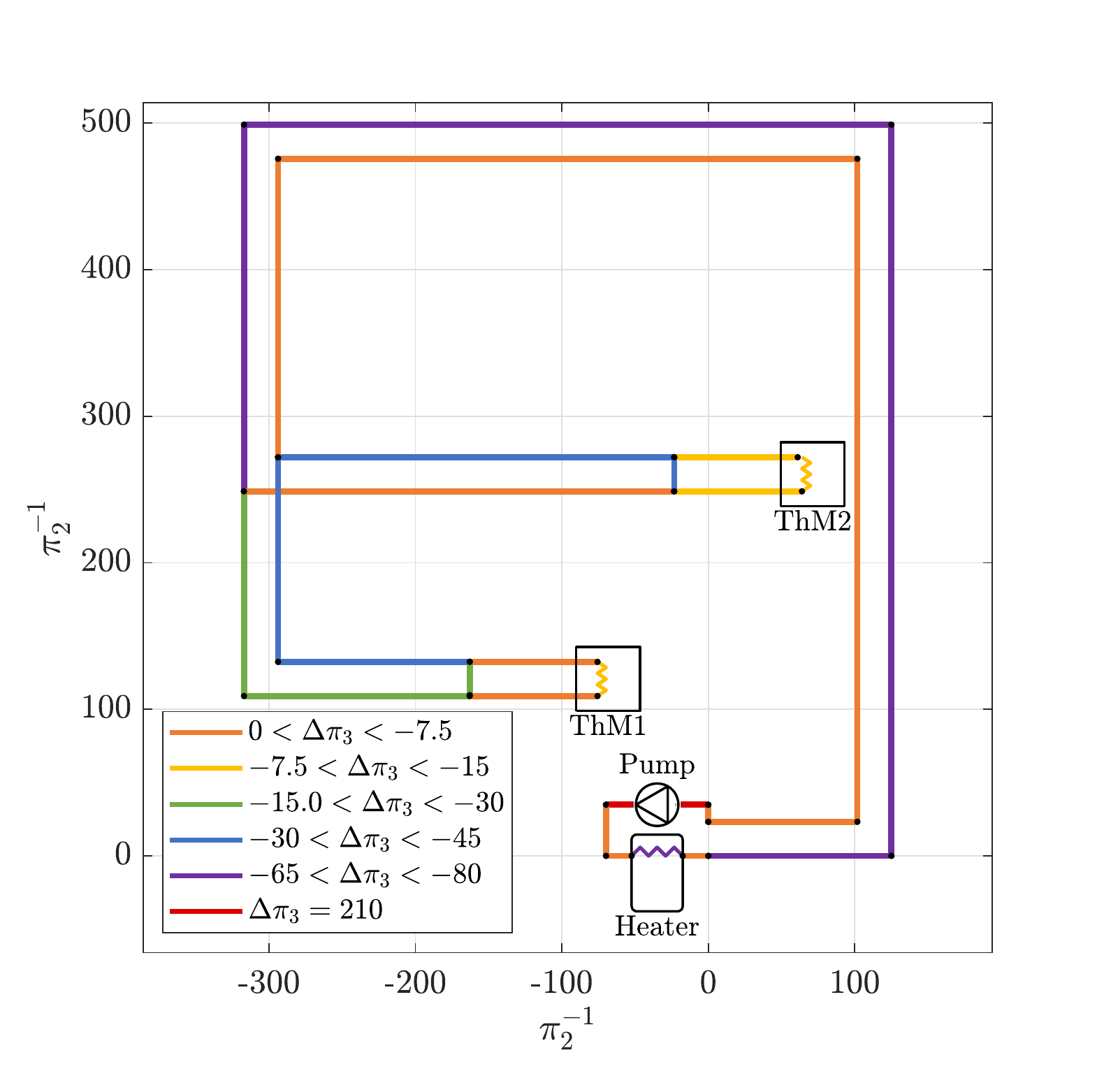}
    \caption{Nondimensional pressure losses in each pipe segment.}
    \label{fig:pressure}
\end{figure}
\begin{figure}
    \centering
    \includegraphics[width = 3in]{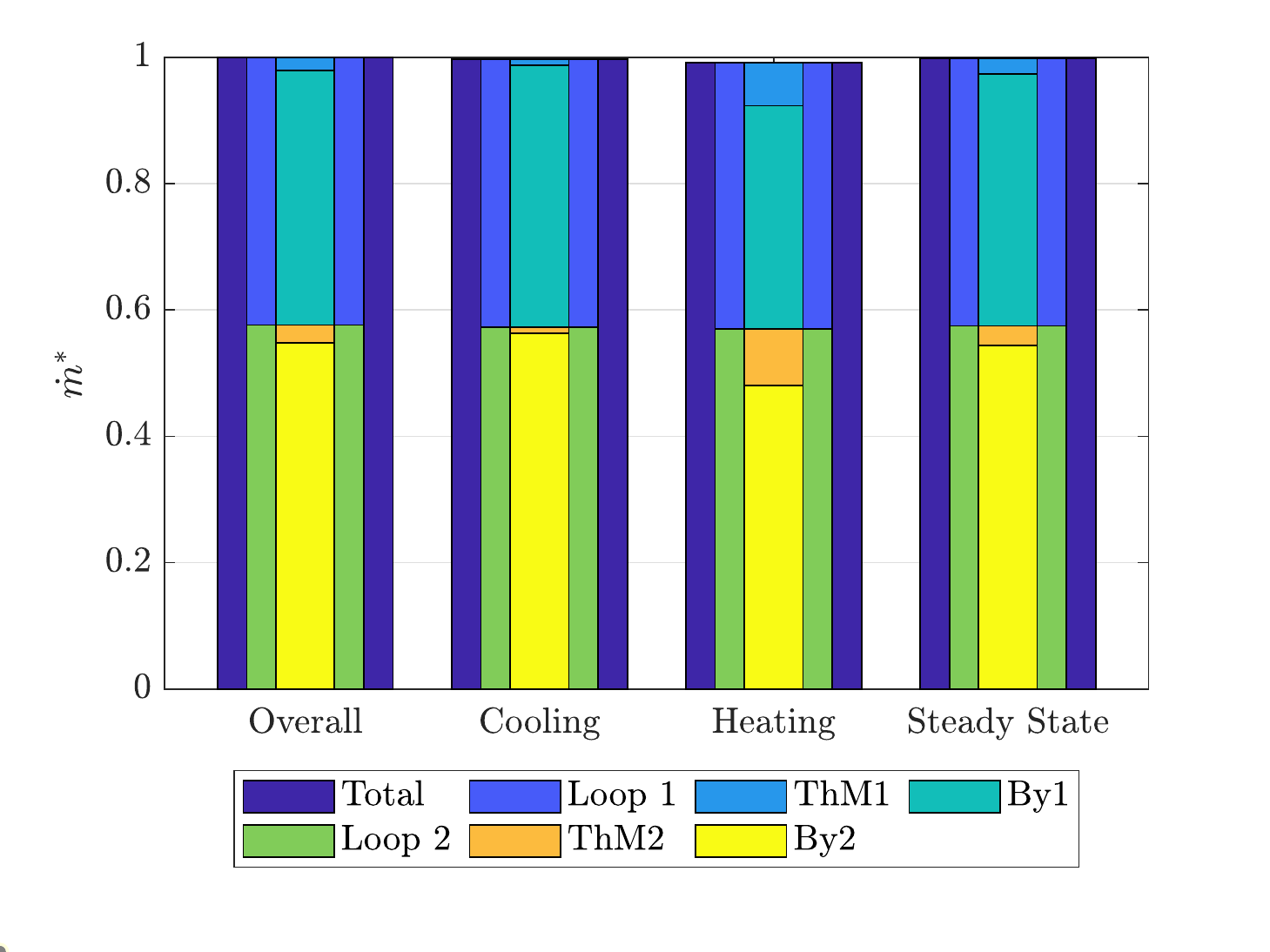}
    \caption{The nondimensional mass flow rate through each pipe segment during different operating periods.}
    \label{fig:massflow}
\end{figure}
\begin{figure}
    \centering
    \includegraphics[width=3in]{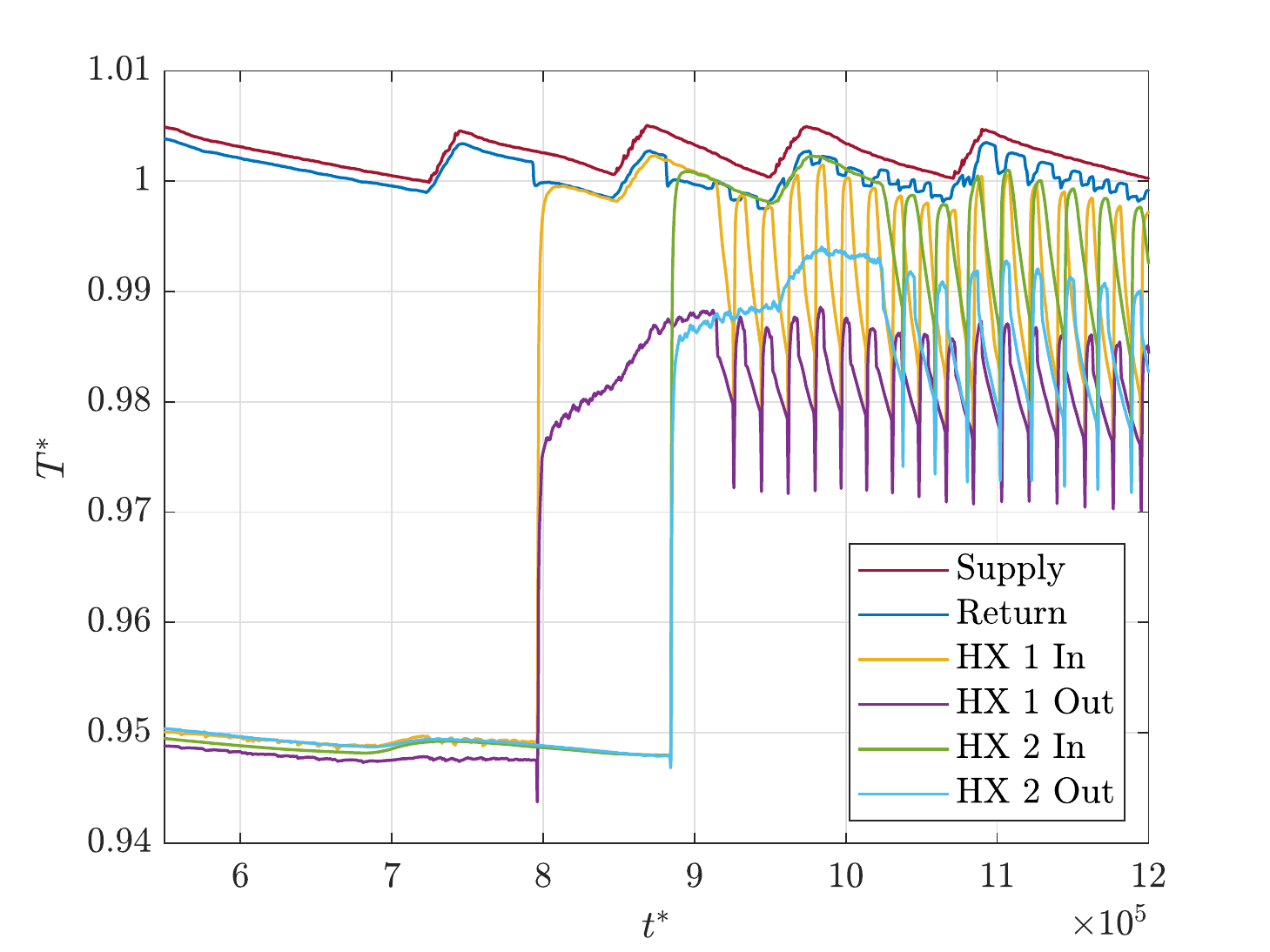}
    \caption{Nondimensional temperature in the pipes during the first heating.}
    \label{fig:pipetemp}
\end{figure}

\begin{table}
\caption{Mean and standard deviation of nondimensional thermal mass temperatures}
\label{table:error}
\centering
\begin{tabular}{lcccc}
\toprule
&\bf{Scale}& \bf{Mean} & \bf{STD}&\bf{Ratio}\\
\midrule
\multirow{2}{*}{$T_{ThM1}$}&Lab& $-9.42\times10^{-3}$ & $10.5\times10^{-3}$&\multirow{2}{*}{7.32}\\
&Full& $-1.29\times10^{-3}$ & $2.15\times10^{-3}$&\\
\midrule
\multirow{2}{*}{$T_{ThM2}$}&Lab& $-6.81\times10^{-3}$ & $8.17\times10^{-3}$&\multirow{2}{*}{2.11}\\
&Full& $-3.22\times10^{-3}$ & $4.07\times10^{-3}$&\\
\bottomrule
\end{tabular}
\end{table}
The response of the pipes in the distribution network is shown in \cref{fig:massflow,fig:pressure,fig:pipetemp}. 
\Cref{fig:pressure} presents the nondimensional pressure losses in the pipes. For reference, the dimensional scale for pressure losses range from 0 to 25 kPa and the pump provides a 75 kPa increase in pressure, which are in the acceptable range for a similar full-scale system. The pressure losses across the heat exhangers are 3.6 and 4.5 kPa for thermal mass 1 and 2 respectively. \Cref{fig:massflow} shows the mass flow split between the network branches. This data is divided into four time intervals, the average mass flow rate, the mass flow rate during the cooling of both thermal masses, the mass flow rate during the heating of both thermal masses, and the mass flow rate during the steady state operation, where the PID controller is working to maintain the current temperature in the thermal masses. The measured values correlate well with relative pressure losses in the network 
. This data is used in the calculation of $\pi_1$ and in the calculation of the heat losses in the network. The values in these plots are validated against the data provided in Ancona et al. \cite{anconaApplicationDifferentModeling2019}.\par
The supply temperature, return temperature, and select pipe temperatures throughout the network are presented in \cref{fig:pipetemp} for the portion of the experiment where both thermal masses are being re-heated for the first time and then maintained at the set temperature. This information was used to validate the temperature dynamics in the pipes. The difference between the supply and return temperature was used to calculate the energy losses throughout the network. Additionally, there is a time delay between changes in the supply temperature and those changes being seen in the return temperature. This delay is caused by the time taken for the operating fluid to circulate through the network. As the fluid dynamics of the system are much faster than the temperature dynamics, this delay can be modeled as an algebraic offset. The magnitude of the algebraic offset seen in the lab-scale data was validated against the data collected from a CFD-based analysis of a DHN’s distribution network presented in Zhao et al. \cite{zhaoInfluencingParametersAnalysis2019}. This paper used the peak-valley method to quantify the delay. In nondimensional time, the average delay in temperature peaks and valleys for users a similar nondimensional distance away from the supply is approximately 4,500, or about 90 seconds in the lab-scale system, similar to the results seen in the lab-scale system. This time delay can have a large impact on the energy consumption of a DHN and can consequently greatly effect the efficiency of a predictive controller. Therefore, accurately capturing this delay in the lab-scale system is critical for control design.\par
\begin{figure*}
\begin{minipage}[l]{.99\columnwidth}
    \centering
    \includegraphics[height=.85in]{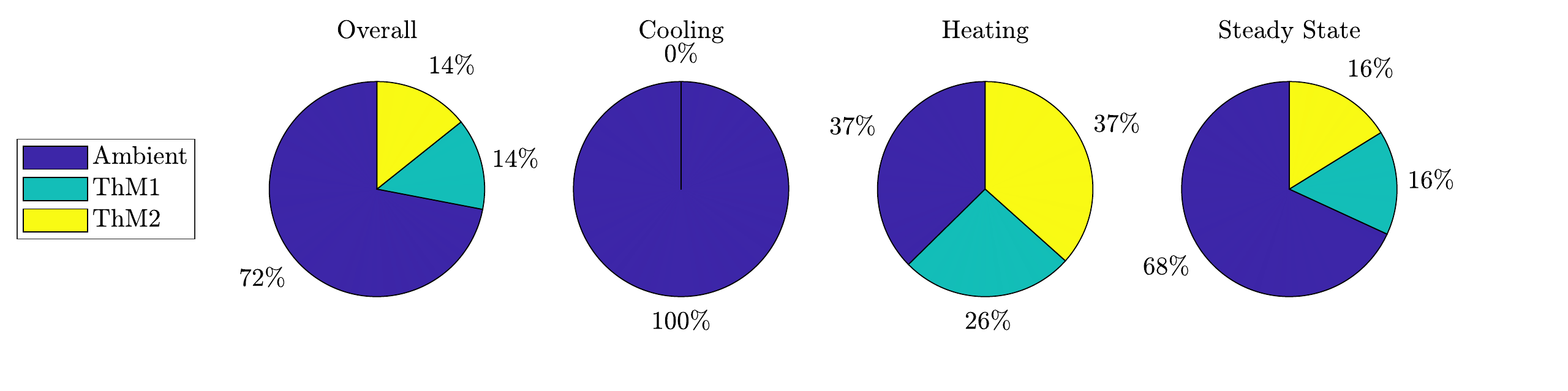}
    \caption{Breakdown of energy leaving the pipe network, separated into heat lost to the ambient, and heat transferred to the two thermal masses during different operating periods.}
    \label{fig:pipeloss}
\end{minipage}\hspace{\columnsep}
\begin{minipage}[r]{.99\columnwidth}
    \centering
    \includegraphics[height=.85in]{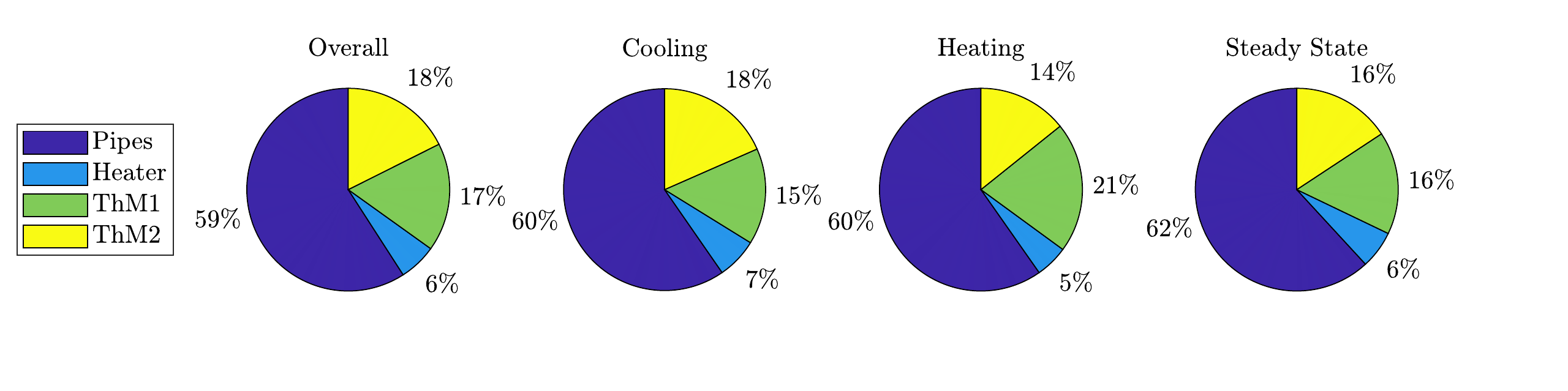}
    \caption{Breakdown of the energy losses in the network, separated into the energy lost by pipe network, the water heater and the two buildings.}
    \label{fig:networkloss}
    \end{minipage}
\end{figure*}

The ratio of heat being lost to the environment compared to the heat being extracted by the heat exchangers is presented in \cref{fig:pipeloss}. The rate of heat lost can be found using the change in enthalpy of the circulating water, calculated according to
\begin{equation}
\label{eq:losstot}
    \dot{Q}_{tot} = \dot{m}_Ic_p\left(T_{s}-T_{r}\right)
\end{equation}
\begin{equation}
\label{eq:lossThM}
    \dot{Q}_{ThM_{i}} = \dot{m}_{HX_{i}}c_p\left(T_{HX_{i} in}-T_{HX_{i} out}\right)
\end{equation}
\begin{equation}
\label{eq:lossdiff}
    \dot{Q}_{tot} = \dot{Q}_{amb}+\dot{Q}_{ThM_{1}}+\dot{Q}_{ThM_{2}}
\end{equation}
where $T_s$ is collected by TM1, $T_r$ by TM2, $T_{HX_{i} in}$ by TM10 and TM5, and $T_{HX_{i} out}$ by TM12 and TM17 respectively.\par
This efficiency measure is divided into the same time intervals as the mass flow rates (overall, cooling, heating, and steady state). The reported average efficiency of a full-scale district heating network is around 70\% \cite{howardMethodologyEvaluationDistrict2020}, while the overall lab-scale is only around 28\% efficient. There are many explanations for the discrepancy between these values. The load diversity in a DHN with many users is much higher than for just the two users being represented here, leading to a higher percentage of the network's operation being in the heating phase, where the lab-scale DHN performs comparably to a full-scale DHN.  Additionally, the size of the thermal masses can also impact this metric, as smaller residential buildings require less heat than larger industrial buildings, thereby decreasing the time spent in the heating phase of operation. Furthermore, in this experiment, no effort was made to reduce the mass flow rate during the times when the buildings were not being heated. A more optimal control algorithm would have reduced the mass flow rate during the cooling periods, reducing the energy wasted and increasing efficiency. Finally, the full-scale efficiency is calculated for an entire year of operation, and heat losses vary drastically depending on the season and ambient temperature.\par
The breakdown of the total heat lost to the environment by each of the components of the system is presented in \cref{fig:networkloss}. Most of of the heat is being lost by pipes, followed by the two thermal masses. The heat lost by the thermal masses includes the heat lost through natural convection and the heat lost due to the cooling of the Peltier junctions. The heat lost by the heater is the smallest source of energy losses in the network. The relative energy loss between the components is fairly consistent between the different phases of operation.

\section{Conclusion}
\label{sec:conclusion}
This paper describes the design and validation of a dynamically similar lab-scale DHN. The equations used to model the system are presented, along with the dimensionless groups that describe the relevant dynamics of the system. Then, representative values for all components of a full-scale DHN based on current literature are provided, along with the corresponding desired lab-scale values. Finally, the scaled DHN is validated by recreating a simulated two day period of operation and compares the results to desired values. The data is presented in the nondimensional form to allow direct comparison to the full-scale data. Future work will use the data collected from the lab-scale DHN to validate modeling techniques developed for use in the design of novel control algorithms. 
\bibliographystyle{elsarticle-num} 
\bibliography{sources}
\end{document}